\documentstyle[stwol,epsfig]{article}
\newcommand{\ccaption}[2]{
  \begin{center}
    \parbox{0.4\textwidth}{
      \caption[#1]{\small\it {#2}}}
  \end{center}    }
\def    \be             {\begin{equation}}
\def    \ee             {\end{equation}} 
\def    \ba             {\begin{eqnarray}}
\def    \ea             {\end{eqnarray}} 

\def    \=              {\;=\;} 
\def    \frac           #1#2{{#1 \over #2}}

\def    \ie             {{\em i.e.\/} }

\def \zprime            {\mbox{$Z^{\prime}$}}
\def \mzprime           {\mbox{$M_{Z^{\prime}}$}}
\def \as   {\mbox{$\alpha_s$}}

\def \et   {\mbox{$E_T$}}                
                
\def \mjj  {\mbox{$m_{jj}$}}                
\newcommand\sss{\scriptscriptstyle}

\def \to   {\mbox{$\rightarrow$}}
\def \GeV    {\mbox{GeV}}

\begin{document}
\begin{titlepage}
\nopagebreak
{\flushright{
        \begin{minipage}{4cm}
        CERN-TH/96-309\\
        hep-ph/9610469\\
        \end{minipage}        }

}
\vfill                        
\begin{center}
{\LARGE { \bf \sc The Fate of the Leptophobic $Z^\prime$}}
\footnote{To appear in the                          
Proceedings of the International Conference on High Energy Physics, 
Warsaw, Poland, July 24--31, 1996.}
\vfill                                
\vskip .5cm                               
{\bf Michelangelo L. MANGANO
\footnote{Presenting author. On leave of absence from INFN, Pisa, Italy.},
Guido ALTARELLI}
\vskip .3cm                                      
{CERN, TH Division, Geneva, Switzerland}\\
\vskip .5cm                               
{\bf Nicola Di Bartolomeo}
\vskip .3cm                                      
{SISSA, Trieste, Italy}\\
\vskip .5cm                               
{\bf Ferruccio FERUGLIO}
\vskip .3cm                                      
{Dipartimento di Fisica dell'Universit\`{a} and INFN, Padova, Italy}\\
\vskip .5cm                               
{\bf Raoul GATTO}
\vskip .3cm                                      
{D\'{e}partement de Physique Th\'{e}orique, Univ. de Gen\`{e}ve,
 Geneva, Switzerland}\\
\end{center}                                                      
\nopagebreak
\vfill                                
\begin{abstract}
We review the main features of the leptophobic-\zprime\ phenomenology,
commenting on the prospects of these models after the recent experimental
results on $R_c$, $R_b$ and after the recent theoretical analyses of jet
production at the Tevatron.
\end{abstract}                                                               
\vskip 1cm
CERN-TH/96-309 \hfill \\
October 1996 \hfill
\vfill       
\end{titlepage}
                             
\section{Introduction}
The possible existence of a new gauge interaction, mediated by a neutral,
massive vector boson (\zprime) and very weakly coupled to
leptons, has
stimulated a large number of studies over the past year\footnote{The number of
papers appeared and the number of issues discussed are so large that in no way
I will be able to quote and discuss them all, given the shortness of this
contribution.}. Although the presence 
of additional $U(1)$ groups is a recurrent feature of models beyond the 
SM, the particular class of theories that we will be interested in here
originated~\cite{Altarelli96,Chiappetta96} from the attempts to explain a very
specific set of anomalies present in recent experimental data from LEP and from
the Tevatron:
\begin{enumerate}
\item a $-2.5\sigma$ discrepancy between the measured and expected value of
$R_c$ (the fraction of hadronic $Z$ decays into charm-quark pairs)\cite{lepewk},
\item a $3.5\sigma$ discrepancy between the measured and expected value of     
$R_b$ (the fraction of hadronic $Z$ decays into bottom-quark
pairs)\cite{lepewk},  and
\item a large discrepancy between the measured and expected rate of high-\et\
jets produced at the Tevatron $p\bar p$ collider\cite{cdfjet}.
\end{enumerate}                                              
The total disappearance of the first effect, the existence of a new precise
measurement of $R_b$ which indicates a sharp decrease of the relative
anomaly~\cite{Warsaw},
and the reduced significance of the high-\et\ jet anomaly due to a better
estimate of the gluon-density systematics~\cite{Tung96}, 
remove completely the need to invoke
such a departure from the SM. Or, to say the least, make it much less appealing
than before. In this presentation I will nevertheless 
dutifully comply to the request of the
session organizers, and present a short review of the main features of the
leptophobic \zprime\ models and of their possible phenomenological
applications.
             
\section{The origin}
What made the proposal of a leptophobic \zprime\ appealing was the coincidence
of several, independent, facts. The large size of $\delta R_b$, for example,
made explanations in terms of virtual effects, such as supersymmetry, unlikely.
The anomaly in $R_c$
would also not easily be understood in a supersymmetric model,
requiring additional features not present in the standard SUSY realizations.
Among possible new tree-level phenomena, the existence of 
an extra $U(1)$, weakly coupled to leptons but sufficiently coupled
to quarks so as to affect the relative rate of $Z$ decays to different quark
flavours, seemed a natural explanation to the $R_c$ and $R_b$ anomalies. Such a
model, however, would have required a fine tuning of the \zprime\ couplings to
different quark flavours, in order to explain the precise agreement between the
measured total $Z$ hadronic width ($\Gamma_h$) and its SM value. 
Such a fine tuning would
have  spoiled the elegance of the model.  A third feature of the data allowed
an elegant solution to the fine-tuning problem: the 1995 values of $R_b$ and
$R_c$ led in fact to the remarkable numerical coincidence that $3\delta
R_b+2\delta R_c=-0.0047\pm 0.0134$, \ie a number compatible with zero. This
can be naturally accomodated by     
assuming a family-independent coupling of the \zprime\ to
up- and down-type quarks. This reduces the number of independent
couplings to quarks from 9 to 3, and makes it easier to enforce
the stability of $\Gamma_h$.
It is this feature that in my
view made the models of ref.~\cite{Altarelli96,Chiappetta96} particularly
appealing. A failure of the relation among the $R_c$ and the $R_b$ anomalies
would make this class of models less interesting. Using the latest data, one now
gets $3\delta R_b+2\delta R_c=0.0047\pm 0.0057$, a number still compatible with
0 at the 1-$\sigma$ level.
The current experimental  
situation, in which  the world averaged $R_b$ is a couple of $\sigma$s
away from the SM and the $R_c$ anomaly has vanished, would however be
explained in a more theoretically-rewarding way by invoking a               
supersymmetric interpretation.    
                                                         
\section{The models and their constraints}
A leptophobic \zprime\ model with the features sketched above
is defined by at least 5 parameters: $\mzprime$, 
the $Z$-\zprime\ mixing angle
$\xi$, the coupling to $L$-handed quarks ($x$) and the couplings to up- and
down-type $R$-handed quarks ($y_{u,d}$). To obtain a consistent model one
should also provide a Higgs sector and complete the set of fermions in order
to achieve anomaly cancellation. In addition to this, but not mandatory, one
might want to consider high-energy embeddings of SM$\times U(1)^\prime$ into
GUT or string models~\cite{GUT}. 
The minimal required Higgs sector can be determined by 
calculating the $U(1)^\prime$ charges of the doublet Higgs fields involved in
the couplings to the known fermions. It is easy to see the these charges are
$x-y_u$, $x-y_d$ and 0 for $u$-quark, $d$-quark and lepton mass terms
respectively. Only one Higgs doublet is therefore necessary if $x=y_u=y_d$
(plus a field to break the $U(1)^\prime$ symmetry), two are needed if
either $x=y_u$ or $x=y_d$ or $x-y_u=x-y_d\ne 0$, and three Higgs doublets are
necessary otherwise. The phenomenology of this extended Higgs sector, by
itself, could lead to interesting phenomena and additional features observable
at the Tevatron~\cite{Glashow96}.

The main constraints on the values of the 5 parameters of the models come from
precision EW data. The \zprime\ contribution to 
a generic EW observable $\cal{O}$ can be parametrized as follows:
\ba
\delta {\cal O} &=& A_{\cal O} \delta\rho_M + B_{\cal O}(x,y_u,y_d) \xi \quad
\mbox{, where} \nonumber \\                            
\delta\rho_M &=& \left[ \left(\frac{\mzprime}{M_Z}\right)^2-1 \right]
 \sin^2\xi \nonumber \\
  & \stackrel{ {\sss \mzprime\gg{M_Z}} }{\sim} &
 \left(\frac{\mzprime\xi}{M_Z}\right)^2 \;.
\ea                                                                       
Typical examples are:
\begin{itemize}
\item the total $Z$ hadronic width:
 \be
   \delta\Gamma_h \sim \xi \left(0.52 x + 0.28 y_u -0.21 y_d\right) \;,
 \ee                                                               
 which sets a strong correlation among the values of the three couplings
 because of the per mille accuracy of the agreement between data and SM;
\item the weak charge of the Cesium nucleus:
 \ba
   \delta Q_{W} &\sim& \xi \left[1-
 \left(\frac{M_Z}{\mzprime}\right)^2 \right] \nonumber \\
 &\times & \left(798 x + 376 y_u +422 y_d \right)\;,
 \ea                                             
(experimentally equal to 1.8) which for $\mzprime\gg{M_Z}$ sets an
independent correlation among $x$, $y_u$ and $y_d$ because of the large
coefficients;
\item the $Z\to b \bar b$ partial hadronic width ($R_b$):
 \be                                                 
   \delta R_b
   \sim \xi\left( -3.2x + 0.7y_u+0.3y_d \right) \; .
 \ee                                          
\end{itemize}
Fits performed using pre-Warsaw data, \ie data incorporating the spring
results, give (for \mzprime=1~TeV, $m_H=300$~GeV and \as=0.118):
\ba                                                           
 & \xi=(2.8 {+0.9\atop-1.3}) \times 10^{-3} &\quad
  x =-1.4  {+0.6\atop-1.4}  \\               
 & y_u= 3.3 {+2.9\atop -1.3}   &\quad
 y_d =1.8 {+2.0\atop -1.0}      
\ea
Fits performed using the Warsaw data~\cite{Warsaw}
give (for \mzprime=1~TeV, $m_H=300$~GeV and \as=0.118):
\ba
& 
 \xi=(2.2 {+0.9\atop-5.3}) \times 10^{-3} &\quad
 x = -0.49 \pm 0.6                  \\     
&
 y_u = 2.0 \pm 1.4                    &\quad
 y_d = 2.1 \pm 1.7                          
\ea 
Notice that, neglecting correlations in the error matrix, 
all couplings and mixing are now individually compatible with 0 to within 1.5
$\sigma$.                                             
                                                              
\section{\zprime\ phenomenology at the Tevatron}
\begin{figure}
\centerline{\epsfig{figure=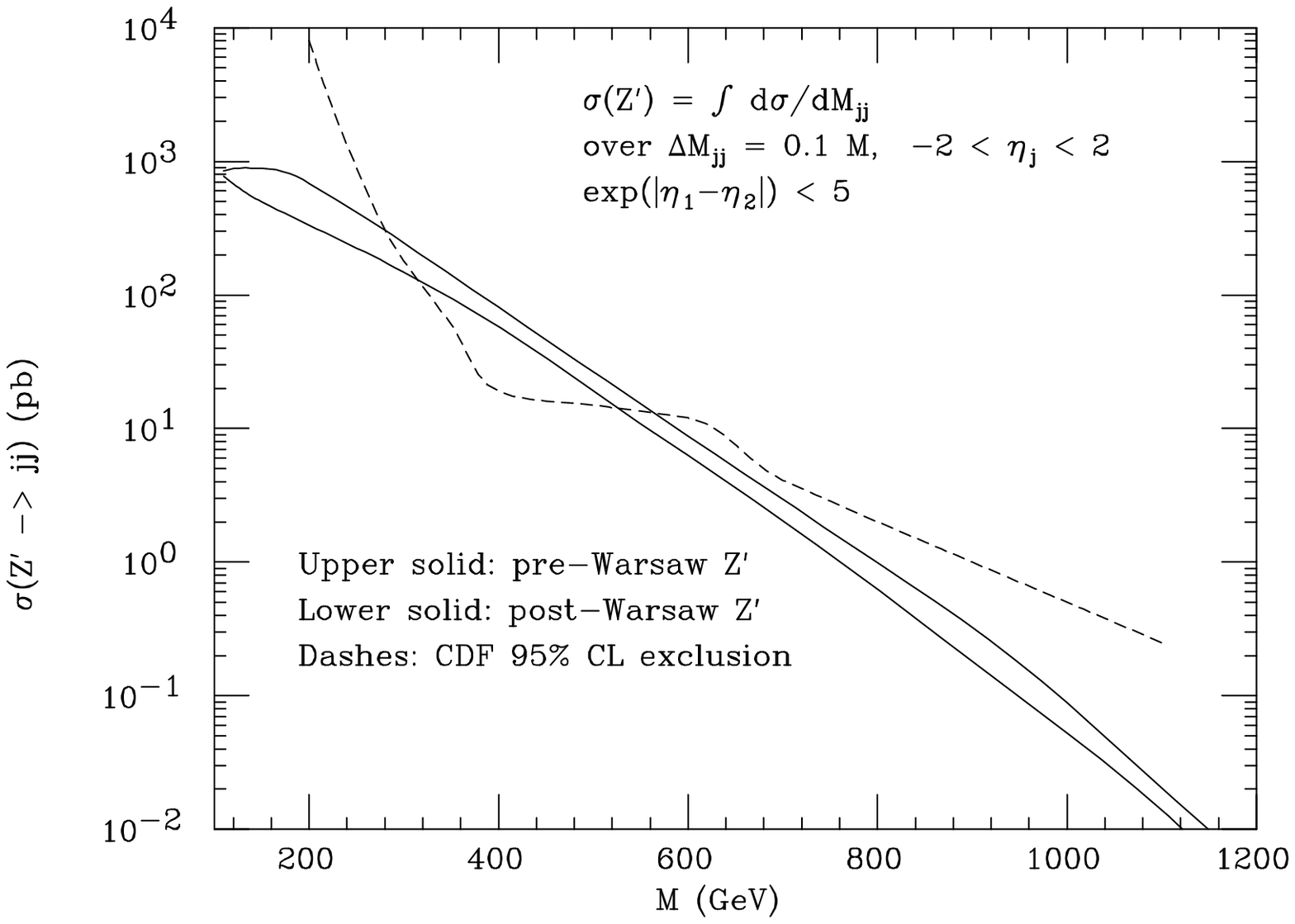,width=0.45\textwidth,clip=}}
\ccaption{}{ \label{cdfmjj}         
Solid: dijet cross section at the Tevatron from the production of a \zprime,
integrated over the mass and rapidity ranges indicated in the plot.
Dashes: 95\%CL Limits on the production cross section of a resonance decaying
into jet
pairs, as a function of the resonance mass, from CDF data.}
\end{figure}
\begin{figure}
\centerline{\epsfig{figure=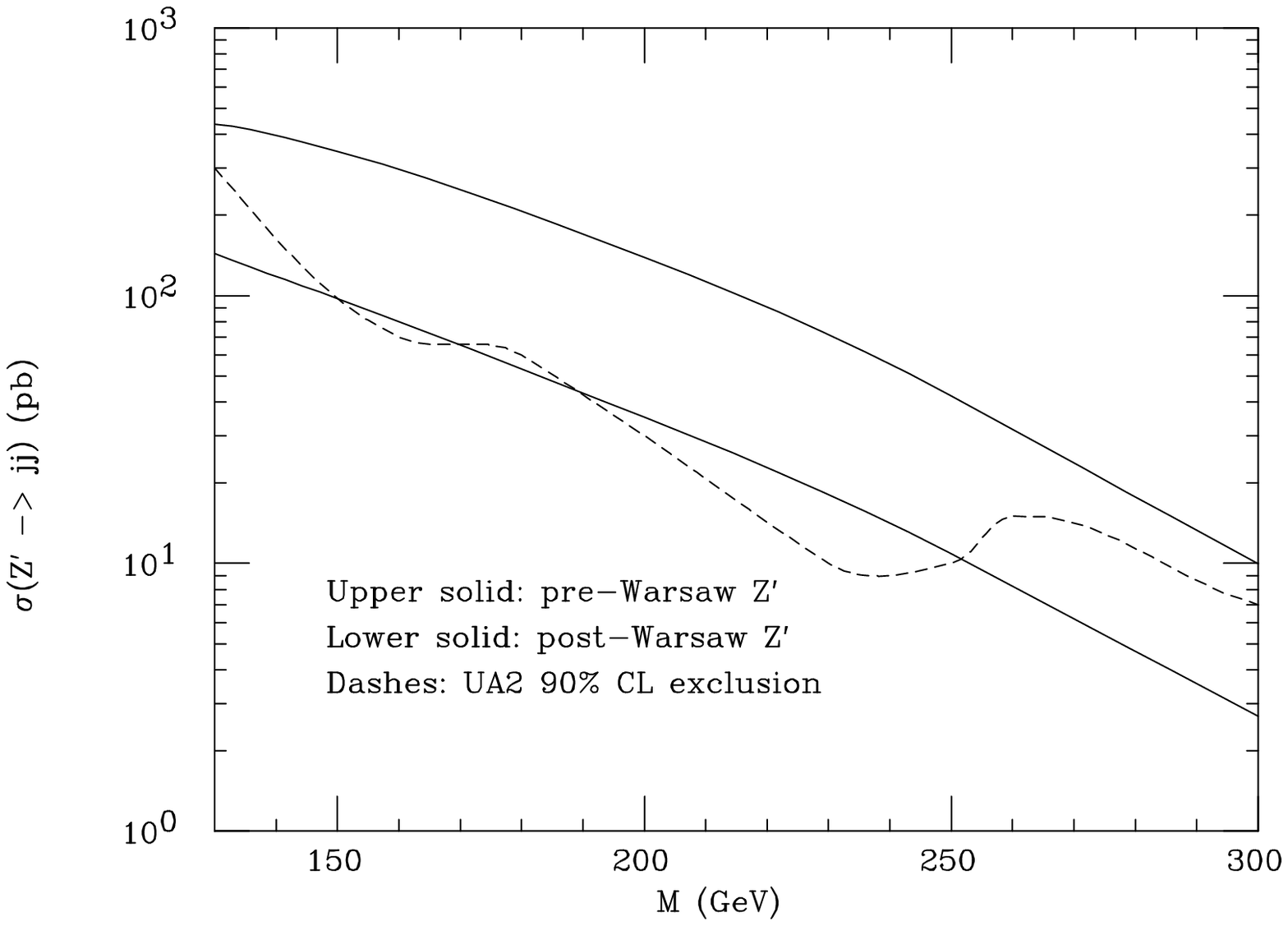,width=0.45\textwidth,clip=}}
\ccaption{}{ \label{ua2mjj}         
Same as fig.~1, from UA2 data at the 90\%CL.}
\vskip 0.2cm
\centerline{\epsfig{figure=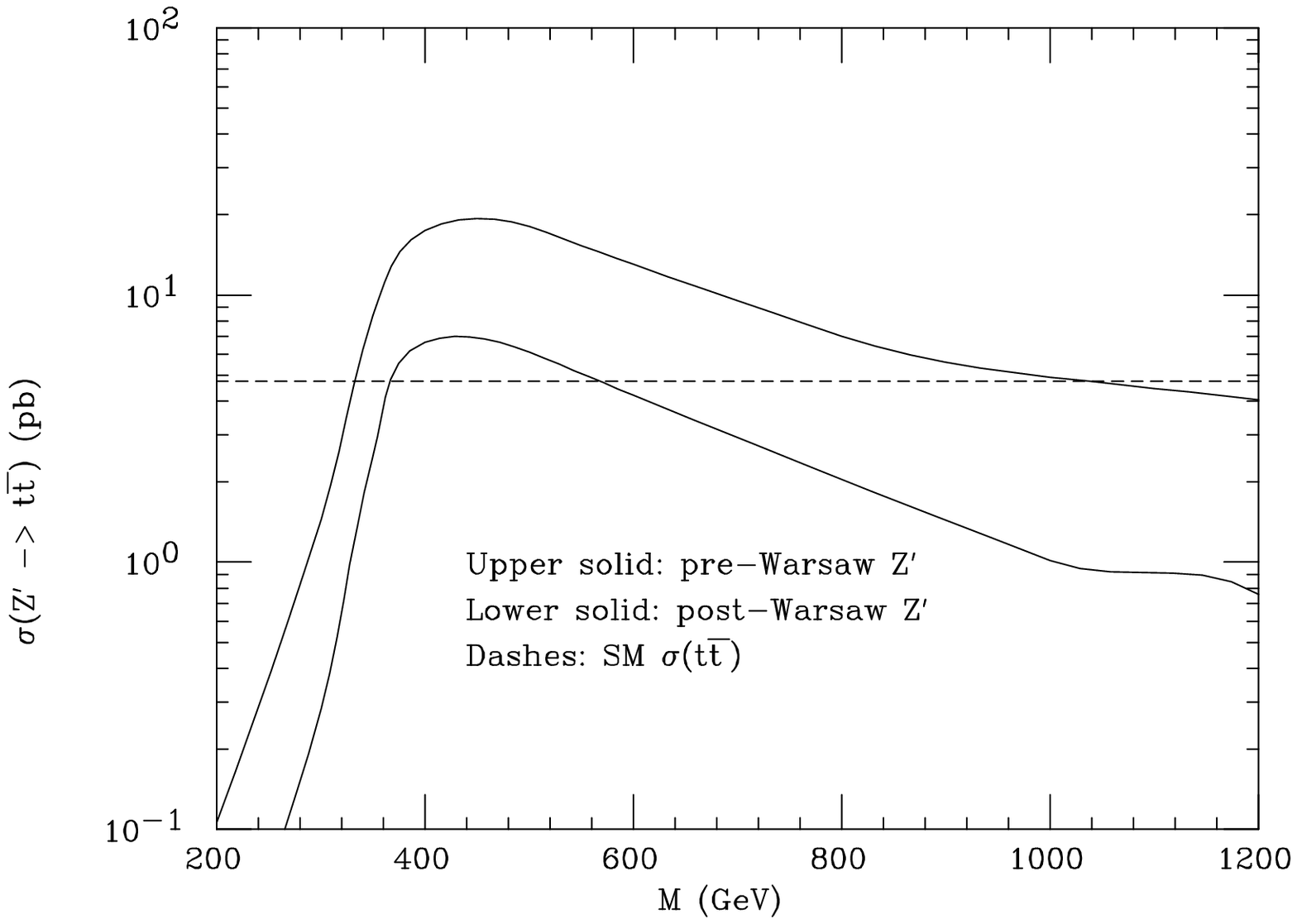,width=0.45\textwidth,clip=}}
\ccaption{}{ \label{cdftop}         
Contribution of the \zprime\ to the total $t\bar t$ cross section at the      
Tevatron (solid), compared to the SM expectation (dashes).}
\vskip 0.2cm
\centerline{\epsfig{figure=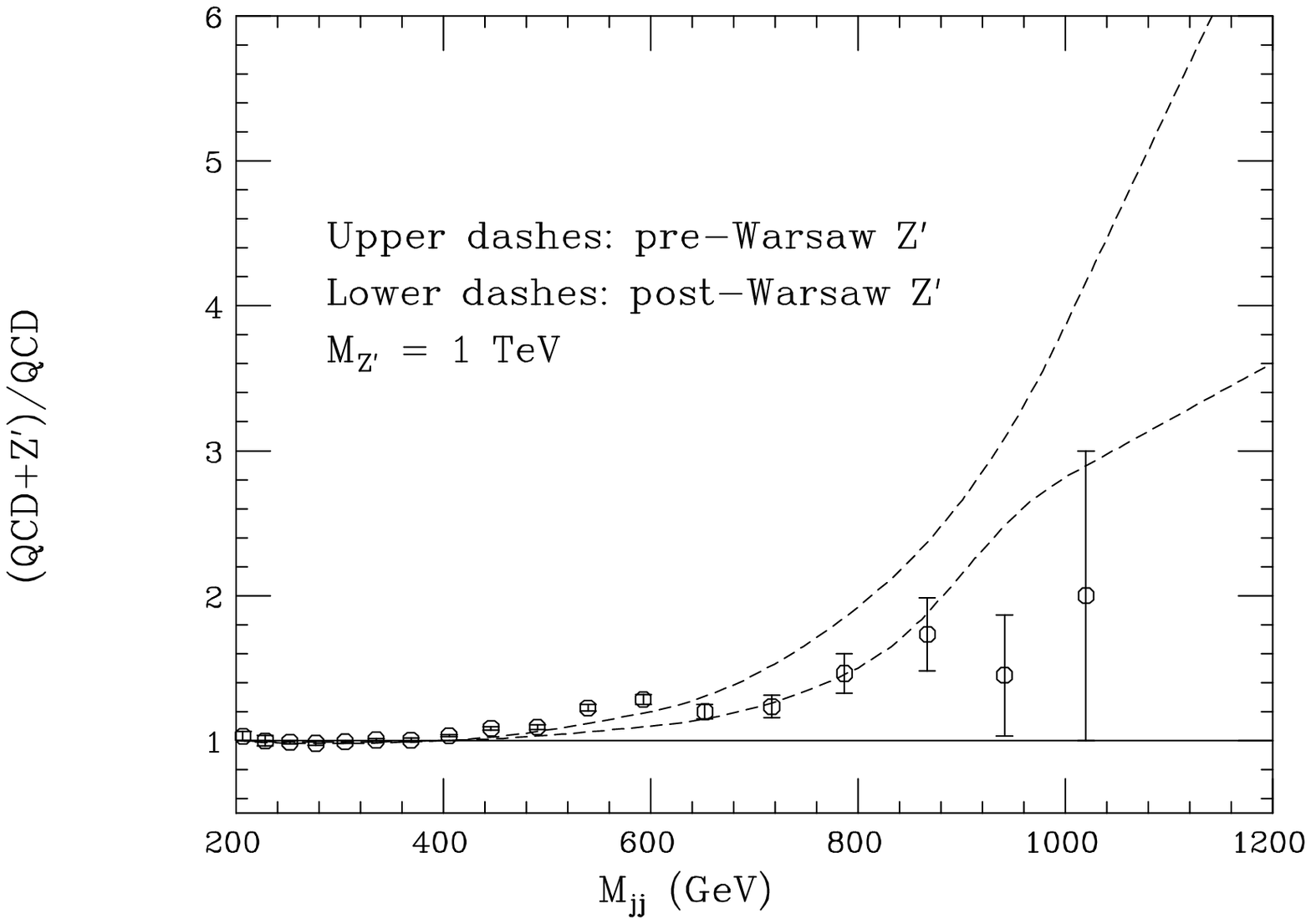,width=0.45\textwidth,clip=}}
\ccaption{}{ \label{cdfrat}         
The standard comparison between data and QCD (or \zprime\ vs QCD) for the
dijet mass spectrum, on a linear                 
scale. We show results for a 1~TeV \zprime, with the set or pre- and
post-Warsaw fitted couplings.}      
\end{figure}

                                                     
As already mentioned, one of the appealing features of the leptophobic \zprime\ 
is its possible connection with the high-\et\ jet rate anomaly at the Tevatron. 
Several other possible implications for the Tevatron physics have been
considered in the literature: the enhancement of the top-quark cross
section~\cite{Stirling96}, the associated production of a light \zprime\ and EW
gauge bosons~\cite{Barger96}, the associated production of $W/Z$ with
neutral and charged Higgs bosons~\cite{Glashow96}, the decay of the \zprime\
into exotic fermions~\cite{Rosner96}, the impact of \zprime-exchange on dijet
angular correlations~\cite{Heyssler96}. 
We present here the impact on some observables which cover the \zprime\ mass
range $130 < \mzprime(\GeV) < 1200$. We used the central values of the fit
results presented in the previous section for both the pre-Warsaw and
post-Warsaw experimental data.
Fig.~\ref{cdfmjj} shows the dijet invariant mass spectrum, in the range
150--1000~GeV, compared to CDF limits on a resonance of mass $M$ and
approximate width $\Gamma\sim 0.1 M$. 
The latest $R_{b,c}$ results reduce the window in which 
a \zprime\ can be excluded from the 280--560~GeV range to the 320--500~GeV
range. The reduction in excluded range is not dramatic as one might expect, due
to decreased width of the \zprime, which partly compensates the loss in total
production rate by making the signal more peaked.
Above 600~GeV the \zprime\ becomes very wide. Only a small fraction of its
rate can be found in a mass region of $\pm 0.1 \mzprime$, so that no limit
can be obtained from bump searches in this region.
Coverage down to lower mass values can be obtained from the
old UA2 analysis~\cite{UA2} (see fig.~\ref{ua2mjj}).
In this case the 90\%CL excluded range is reduced to a window between 200 and
250 GeV.
The effect of a \zprime\ on the top production cross section, compared to the
SM expectation~\cite{Catani96}, is shown in fig.~\ref{cdftop}. This effect used
to set
the strongest constraints on a \zprime\ with mass in the region between 300 and
1000~GeV. Now one can exclude only the region 350--600 GeV, similar to the
window excluded by the CDF searches in the \mjj\ spectrum.

The effect of the \zprime\ on the dijet mass spectrum at the Tevatron, compared
to current CDF data, is shown in fig.~\ref{cdfrat}, for 1~TeV \zprime.
As already pointed out in
ref.~\cite{Altarelli96}, the central values of the typical pre-Warsaw fit would
give too large a jet rate in the region around 1000~GeV, 
unless the \zprime\ mass
were larger than 1.2~TeV. The reduction in couplings due to the latest fit
improves a bit the situation, but again less than naively expected, due to the
reduced width which reduces the dijet mass smearing. One could nevertheless
argue that the current fits produce a reasonable agrement with the high-mass
behaviour of the data. If one 
neglected the indications~\cite{Tung96} that the uncertainty in the gluon
density could reduce the jet anomaly, one should accept this as the
only remaining evidence for the possible existence of a leptophobic \zprime.
                                                        
\section{Conclusions}
The new measurements of $R_{c,b}$ seriously undermine the phenomenological
motivations for the class of leptophobic \zprime\ models recently considered
in the literature. 
Due to the weakening of the fitted couplings (which are now consistent with 0
at the 1.5-$\sigma$ level) the mass regions in which a \zprime\ would have
given a signal in hadronic collisions are significantly reduced. A \zprime\ in
the range below 200 GeV and above 600 GeV  would have easily escaped detection
so far.  The high-\et\ jet anomaly at CDF remains as the only set of data
supporting, but not necessarily mandating, the existence of a \zprime.

\end{document}